\documentstyle[12pt]{article}

\begin{document}

\begin{flushright}
IMSc/2019/09/07 
\end{flushright} 

\vspace{2mm}

\vspace{2ex}

\begin{center}

{\large \bf LQ Stars : Modifying the}

\vspace{4ex}

{\large \bf Tolman -- Oppenheimer -- Volkoff equations a la LQC}

\vspace{8ex}

{\large  S. Kalyana Rama}

\vspace{3ex}

Institute of Mathematical Sciences, HBNI, C. I. T. Campus, 

\vspace{1ex}

Tharamani, CHENNAI 600 113, India. 

\vspace{2ex}

email: krama@imsc.res.in \\ 

\end{center}

\vspace{6ex}

\centerline{ABSTRACT}

\begin{quote} 

We rewrite the Tolman -- Oppenheimer -- Volkoff (TOV) equations
for four and higher dimensional static spherically symmetric
stars so that they resemble the equations for anisotropic
cosmology.  This becomes possible by treating the curvature
terms of the sphere as part of the matter sector. Comparing then
with the effective equations in the Loop Quantum Cosmology --
inspired models, we propose analogous modifications to the TOV
equations which contain one arbitrary function. A linear
function gives the TOV equations. If the function and all its
derivatives are finite then the solutions are non singular. For
$sin \; x \;$, an `image star' appears beyond the central region
of maximal pressures in the original star. This is evocative of
a bouncing universe in Loop Quantum Cosmology. We discuss
possible observational consequences of these features.

\end{quote}

\vspace{2ex}


%
%
%
%
%
%
%

\newpage 

\vspace{4ex}

\begin{center}

{\bf 1. Introduction}  

\end{center}

\vspace{2ex} 

General relativity equations for cosmology lead to big bang
singularities where the pressures and the densities diverge and
the size of the universe vanishes. These singularities are
resolved in Loop Quantum Cosmology (LQC) which is obtained from
Loop Quantum Gravity (LQG). In LQC, due to quantum effects, the
densities and the pressures in the universe remain non
divergent, the size of the universe remains non vanishing, and
the universe `bounces back' as one goes back in time. The
quantum dynamics in LQC can be well described by effective
equations which lead to general relativity equations in the
classical limit \cite{b} -- \cite{as}. Using the higher
dimensional formulation of LQG given in \cite{btt, btt2, btt3},
similar effective equations can also be obtained for higher
dimensional cosmology \cite{z, rs}.

Recently, we constructed LQC -- inspired models by empirically
generalising the effective equations in LQC from four to $D$
dimensions, and also by including arbitrary functions
\cite{k16}. In what is referred to as $\bar{\mu}-$scheme, the
LQC -- inspired models contain one arbitrary function $f(x)$
with the only requirement that it $\to x$ as $x \to 0 \;$. By
construction, general relativity equations follow for $f(x) = x$
and the effective equations in LQC follow for $D = 4$ and $f(x)
= sin \; x \;$. We then studied the variety of evolutions
possible in these models for different choices of $f (x) \;$,
their possible relations with Brans -- Dicke theories, and their
application to M theory cosmology \cite{k17, k18, k19}.

Consider the Tolman -- Oppenheimer -- Volkoff (TOV) equations
which are general relativity equations for four dimensional
static spherically symmetric stars. Consider too stars in $D =
(n_c + n + 2)$ dimensional spacetime, with $n \ge 2 \;$, which
are suitably smeared over the $n_c$ dimensional compact space
and are static and spherically symmetric in the $(n + 1)$
dimensional non compact transverse space. Stars in string or M
theory may be studied in this setting \cite{k13, k15}. For the
sake of brevity, in the following, we may refer to the general
relativity equations for such higher dimensional stars also as
TOV equations.

As for cosmology, it is of great interest to obtain modified
equations for stars also which will resolve the singularities
that may arise, and will lead to TOV equations in the classical
limit. The modified equations may then be used to study, for
example, the static spherically symmetric stars and their
stability properties where, now, the singularities may be
resolved. Such modified TOV equations are perhaps derivable from
LQG but, to our knowledge, they are not known at present.

In this paper, we propose to modify the TOV equations for stars.
The modified equations are similar to the effective equations
for cosmology obtained using LQC ; they contain one arbitrary
function $f(x) \;$; and, by construction, lead to the TOV
equations for $f(x) = x \;$.

For this purpose, we first consider the TOV equations for $(n_c
+ n + 2)$ dimensional stars. We realise that, by treating the
curvature terms of the $n$ dimensional sphere as part of the
matter sector, the TOV equations can be rewritten so that they
resemble the general relativity equations for anisotropic
cosmology. We then consider the effective equations in LQC and
in the LQC -- inspired models. An inspection of the similarities
between the general relativity equations for stars and for
cosmology, and the structure of the quantum effective equations
for cosmology obtained using LQC, then immediately suggests the
analogous modifications to the TOV equations. The modified
equations contain one arbitrary function $f(x) \;$. By
construction, the choice $f(x) = x$ gives general relativity
equations; another canonical choice is $f(x) = sin \; x$ as in
LQC. Other choices of $f(x) \;$, as in \cite{k17} for example,
are also possible.

TOV equations can be solved explicitly for an $(n + 2)$
dimensional star of constant density. In LQC also, where $f(x) =
sin \; x \;$, explicit solutions can be found for an isotropic
universe whose constituents have an equation of state $p =
(const) \; \rho \;$. But we are unable to find explicit
solutions to the modified TOV equations for any non trivial
function $f(x) \;$. Hence, in this paper, we analyse these
equations from a general perspective only.

We find that the solutions to the modified equations, which
decsribe a star, are non singular if $f(x)$ and all its
derivatives are finite : the densities, the pressures, the
curvature invariants, and the inverse of the areal radius all
remain finite and non divergent. Thus, the stars of constant
density will have non divergent pressures at their centers.
Therefore they will be non singular, and their compactness
factors can be close to but less than the Buchdahl bound.

We find that the choice $f(x) = sin \; x \;$ results in novel
and unique features which may be observable. One of these
features is the appearance of an `image star' beyond the central
region of maximal pressures in the original star, this being
similar to the bouncing phase of an anisotropic universe in LQC
where there is an `image universe' in the past of the bounce.
The term Loop Quantum (LQ) stars, used in the title of the
paper, perhaps describes such objects concisely and is evocative
of LQC's bouncing universes.

This paper is organised as follows. In section {\bf 2}, we
present the TOV equations for $(n_c + n + 2)$ dimensional stars.
We include the curvature terms of the $n$ dimensional sphere
among the matter sector, and rewrite the TOV equations so as to
resemble the equations for anisotropic cosmology. In section
{\bf 3}, we modify the TOV equations for $(n_c + n + 2)$
dimensional stars, and then write them in detail for $(n + 2)$
dimensional stars. In section {\bf 4}, we analyse the non
singularity of the resulting solutions and describe the novel
features of the modified stars. In section {\bf 5}, we discuss
possible observational consequences and falsifiable predictions
arising from `image star' for $f(x) = sin \; x \;$. In section
{\bf 6}, we summarise the paper and conclude by mentioning a
couple of topics for further studies. In Appendix {\bf A}, we
present the $(n + 2)$ dimensional TOV equations and their
solution for stars of constant density. In Appendix {\bf B}, for
ease of reference, we write down the general relativity
equations for $D$ dimensional anisotropic cosmology and the
corresponding modified equations in the LQC -- inspired models.


\vspace{4ex}

\centerline{\bf 2. General Set Up for
$\mathbf D = (n_c + n + 2)$ dimensional stars}

\vspace{2ex}

Let the spacetime be $D = (n_c + n + 2)$ dimensional where $n_c
\ge 0$ and $n \ge 2 \;$, let the $n_c$ dimensional space be
toroidal, and let the $(n + 1)$ dimensional transverse space be
non compact. Stars in string or M theory may be studied in this
setting \cite{k13, k15}. Let $x^M = (t, x^i, r, \theta^a)$ be
the coordinates where $i = 1, 2, \cdots, n_c$ and $a = 1, 2,
\cdots, n \;$. The general relativity equations are given, in
the standard notation with $\kappa^2 = 8 \pi G_D \;$, by
\begin{equation}\label{rt}
R_{M N} - \frac{1}{2} \; g_{M N} \; R = \kappa^2 \; T_{M N}
\; \; , \; \; \; \sum_M \nabla_M T^M_{\; \; N} = 0
\end{equation}
where $T_{M N}$ is the energy momentum tensor of the constituent
matter. We will consider stars which are static and spherically
symmetric in the transverse space. The suitable line element $d
s$ is then given by
\begin{equation}\label{ds} 
d s^2 = - \; e^{2 \lambda^0} d t^2 + \sum_i e^{2 \lambda^i}
(d x^i)^2 + e^{2 \lambda} d r^2 + e^{2 \sigma} d \Omega_n^2
\end{equation}
where $d \Omega_n$ is the line element on an $n$ dimensional
unit sphere and the fields depend on $r$ only. Let the non
vanishing components of $T^M_{\; \; N}$ be given by
\[
\left( T^0_{\; \; 0}, \; T^i_{\; \; i}, \; T^r_{\; \; r}, \;
T^a_{\; \; a} \right) = \left( p_0, \; p_i, \; \Pi, \; p_a
\right)
\]
where $p_0 = - \rho$ and $p_a = p$ for all $a \;$. Then, after a
long but straightforward algebra, equations following from
(\ref{rt}) may be written as
\begin{eqnarray}
\Pi_r & = & \sum_\alpha (- \Pi + p_\alpha) \;
\lambda^\alpha_r \label{1} \\
& & \nonumber \\
\sum_{\alpha \beta} G_{\alpha \beta} \; \lambda^\alpha_r \;
\lambda^\beta_r & = & 2 \; \kappa^2 \; \Pi \; e^{2 \lambda}
+ n (n - 1) \; e^{2 \lambda - 2 \sigma} \label{e}
\end{eqnarray}
\begin{equation}\label{lrr}
\lambda^\alpha_{r r} + (\Lambda_r - \lambda_r) \;
\lambda^\alpha_r = \kappa^2 \; \sum_\beta G^{\alpha \beta}
\left( \Pi + p_\beta \right) \; e^{2 \lambda} + \delta^{\alpha
a} \; (n - 1) \; e^{2 \lambda - 2 \sigma}
\end{equation}
where the subscripts $r$ denote $r-$derivatives and we have defined
\begin{eqnarray}
\alpha = (0, i, a) & , & 
\lambda^\alpha = (\lambda^0, \lambda^i, \lambda^a) 
\; \; , \; \; \; 
p_\alpha  = (p_0, p_i, p_a) \nonumber \\
& & \nonumber \\
G_{\alpha \beta} = 1 - \delta_{\alpha \beta} & , & 
G^{\alpha \beta} =  \frac{1}{D - 2} - \delta^{\alpha \beta}
\; \; , \; \; \; 
\Lambda = \sum_\alpha \lambda^\alpha \label{def1}
\end{eqnarray}
with $\lambda^a = \sigma$ and $p_a = p$ for all $a \;$. It
follows from these definitions that 
\begin{eqnarray*}
\sum_\beta G^{\alpha \beta} \; G_{\beta \gamma} =
\delta^\alpha_{\; \; \gamma} & , & \sum_\beta G^{\alpha \beta}
\; = \; \frac{1}{D - 2} \\ 
& & \\
\sum_\beta G^{\alpha \beta} \left( \Pi + p_\beta \right)
& = & - p_\alpha + \frac{T}{D - 2} 
\end{eqnarray*}
where $T = \Pi + \sum_\beta p_\beta \;$. We now make several
remarks and rewrite equations (\ref{1}) -- (\ref{lrr}) in a
suitable form, see equations (\ref{el}), (\ref{1*Il}), and
(\ref{le}) below.

\begin{itemize}

\item

If the constituent matter is assumed to be made up of ${\cal N}$
independent, non interacting components then
\begin{equation}\label{tmni}
T_{M N} = \sum_I T_{M N (I)} \; \; ; \; \; \; 
\sum_M  \nabla_M  T^M_{\; \; N (I)} = 0 
\end{equation}
where $I = 1, 2, \cdots, {\cal N} \;$ and $T_{M N (I)}$ are the
seperately conserved individual energy momentum tensors. Let the
non vanishing components of $T^M_{\; \; N (I)}$ be given by
\[
\left( T^0_{\; \; 0 (I)}, \; T^i_{\; \; i (I)}, \;
T^r_{\; \; r (I)}, \; T^a_{\; \; a (I)} \right) = \left(
p_{0 I}, \; p_{i I}, \; \Pi_I, \; p_{a I} \right)
\]
where $p_{0 I} = - \rho_I$ and $p_{a I} = p_I$ for all $a \;$.
Then equations (\ref{tmni}) give 
\begin{equation}\label{1I}
\Pi = \sum_I \Pi_I \; \; , \; \; \; 
p_\alpha = \sum_I p_{\alpha I} \; \; \; ; \; \; \; \;
(\Pi_I)_r \; = \; \sum_\alpha (- \Pi_I + p_{\alpha I}) \;
\lambda^\alpha_r 
\end{equation}
from which it follows that if $p_{\alpha I} = c_{\alpha I} \Pi_I
\;$, with $c_{\alpha I}$ constants, then
\begin{equation}\label{phiI}
\Pi_I \; = \; \Pi_{I 0} \; 
e^{\phi^I} \; \; , \; \; \; \phi^I \; = \;
\sum_\alpha (c_{\alpha I} - 1) \; \lambda^\alpha 
\end{equation}
where $\Pi_{I 0}$ is a constant.


\item 

Consider a matter component labelled by $I = *$ and for which
$p_{\alpha *}$ are given by 
$p_{a *} = p_*$ for all $a$ and 
\begin{equation}\label{*}
p_{0 *} = p_{i *} = \Pi_* \; \; , \; \; \; 
p_* = \frac {n - 2} {n} \; \Pi_* \; \; . 
\end{equation}
Then one has $T_* = \Pi_* + \sum_\beta p_{\beta *} = (n_c + n)
\; \Pi_* \;$,
\[
\sum_\beta G^{\alpha \beta} \left( \Pi_* + p_{\beta *} \right) =
\Pi_* - p_{\alpha *} = \delta^{\alpha a} \; \frac {2} {n} \;
\Pi_* 
\]
and, using equation (\ref{phiI}), $\Pi_* = \Pi_{* 0} \; e^{- 2
\sigma}$ where $\Pi_{* 0}$ is a constant. It hence follows that
the last terms in equations (\ref{e}) and (\ref{lrr}), which are
the curvature terms of the $n$ dimensional sphere, can be taken
into account by setting $2 \kappa^2 \Pi_{* 0} = n (n - 1)\;$ and
writing
\begin{equation}\label{1*I}
\tilde{\Pi} = \sum_{\tilde{I}} \Pi_{\tilde{I}} \; \; , \; \; \;
\tilde{p}_\alpha = \sum_{\tilde{I}} p_{\alpha \tilde{I}}
\; \; \; ; \; \; \; \; (\Pi_{\tilde{I}})_r \; = \;
\sum_\alpha (- \Pi_{\tilde{I}}
+ p_{\alpha \tilde{I}}) \; \lambda^\alpha_r 
\end{equation}
where $\tilde{I} = (*, \; I) = *, 1, 2, \cdots, {\cal N}
\;$. Thus $\tilde{\Pi} = \Pi_* + \Pi$ and $\tilde{p}_\alpha =
p_{\alpha *} + p_\alpha$ where $\Pi$ and $p_\alpha$ are given by
equation (\ref{1I}).

\item

Changing the
independent variable from $r$ to $l \;$ given by
\begin{equation}\label{rl}
e^\lambda \; d r = d l \; \; , 
\end{equation} 
the line element given in equation (\ref{ds}) becomes 
\begin{equation}\label{dsl} 
d s^2 = - \; e^{2 \lambda^0} d t^2 + \sum_i e^{2 \lambda^i}
(d x^i)^2 + d l^2 + e^{2 \sigma} d \Omega_n^2 
\end{equation}
and, for any function $\psi(r(l))$, one has 
\begin{equation}\label{psil} 
\psi_l = e^{ - \lambda} \; \psi_r \; \; , \; \; \;
\psi_{l l} = e^{- 2 \lambda} \; \left( \psi_{r r}
- \lambda_r \; \psi_r \right)
\end{equation}
where the subscripts $l$ denote $l-$derivatives. Equations
(\ref{e}), (\ref{lrr}), and (\ref{1*I}) then give
\begin{eqnarray}
\sum_{\alpha \beta} G_{\alpha \beta} \; \lambda^\alpha_l \;
\lambda^\beta_l & = & 2 \; \kappa^2 \; \tilde{\Pi} \label{el} \\
& & \nonumber \\
\lambda^\alpha_{l l} + \Lambda_l \; \lambda^\alpha_l & = &
\kappa^2 \; \sum_\beta G^{\alpha \beta}
\left( \tilde{\Pi} + \tilde{p}_\beta \right) \label{lll} \\
& & \nonumber \\ 
\tilde{\Pi} = \sum_{\tilde{I}} \Pi_{\tilde{I}} & , &
\tilde{p}_\alpha = \sum_{\tilde{I}} p_{\alpha \tilde{I}} 
\; \; , \; \; \;
\tilde{I} = *, 1, 2, \cdots, {\cal N} \; \; , \nonumber \\
& & \nonumber \\
(\Pi_{\tilde{I}})_l & = & \sum_\alpha (- \Pi_{\tilde{I}}
+ p_{\alpha \tilde{I}}) \; \lambda^\alpha_l \; \; . 
\label{1*Il} 
\end{eqnarray}

\item 

After solving equations (\ref{el}) -- (\ref{1*Il}) and obtaining
$(\lambda^\alpha, \; \Pi_{\tilde{I}}, \; p_{\alpha \tilde{I}})
\;$ in terms of $l \;$, one needs to find $\lambda (l)$ and
$r(l) \;$, obtain $l(r)$ by a functional inversion, and express
the solutions in terms of $r \;$. However, there is a freedom in
defining the radial coordinate $r \;$. We choose it to be the
areal radius so that
\[
d l^2 + e^{2 \sigma} d \Omega_n^2 \; = \;
e^{2 \lambda} d r^2 + r^2 d \Omega_n^2 \; \; .
\]
Then $r(l) = e^{\sigma(l)} \;$ and, hence, $\lambda(l)$ is given
by
\begin{equation}\label{lal}
e^{- \lambda} \; = \; \frac {d r} {d l} \; = \;
e^\sigma \; \sigma_l \; \; .
\end{equation}

\item

Define $Y_\alpha$ by
\begin{equation}\label{ya}
Y_\alpha = \sum_\beta G_{\alpha \beta} \; \lambda^\beta_l
= \Lambda_l - \lambda^\alpha_l \; \; \; \Longrightarrow \; \; \;
\lambda^\alpha_l = \sum_\beta G^{\alpha \beta} \; Y_\beta
\end{equation}
which, together with equation (\ref{el}), implies that 
\[ 
\sum_\alpha Y_\alpha = (D - 2) \; \Lambda_l \; \; , \; \; \;
\sum_\alpha \lambda^\alpha_l \; Y_\alpha =
2 \; \kappa^2 \; \tilde{\Pi} \; \; .
\]
Then, subtracting equation (\ref{el}) from equation (\ref{lll})
gives
\begin{equation}\label{le}
\lambda^\alpha_{l l} + \sum_\beta \frac {\left( \lambda^\alpha_l
- \lambda^\beta_l \right) \; Y_\beta} {D - 2} \; = \; \kappa^2
\; \sum_\beta G^{\alpha \beta} \; \left( - \tilde{\Pi} +
\tilde{p}_\beta \right) \; \; .
\end{equation}
Note that differentiating both sides of equation (\ref{el}) with
respect to $l$ and using equations (\ref{ya}) and (\ref{le})
immediately gives the conservation equation 
\begin{equation}\label{1*l} 
\tilde{\Pi}_l \; = \; \sum_\alpha (- \tilde{\Pi}
+ \tilde{p}_\alpha) \; \lambda^\alpha_l \; \; .
\end{equation}



\end{itemize}


\vspace{4ex}

\begin{center}

{\bf 3. Modifying the  equations for stars a la LQC}

\end{center}

\vspace{2ex} 

Note the similarities between equations (\ref{el}) and
(\ref{le}) which describe stars, and the equations (\ref{ce})
and (\ref{lce}) given in Appendix {\bf B} which describe
anisotropic cosmology. Both of these sets of equations are
obtained from the general relativity equations (\ref{rt}).

Consider the effective equations in the LQC \cite{as}; and the
equations in the LQC -- inspired models which we had constructed
recently by empirically generalising the effective LQC equations
from four to $D$ dimensions, and also by including arbitrary
functions \cite{k16} -- \cite{k19}. These effective equations
contain one arbitrary function $f(x)$ in what is referred to as
$\bar{\mu}-$scheme, and are given by equations (\ref{fgx}) --
(\ref{e3}) in Appendix {\bf B}. The function $f(x) = x$ for
general relativity and $= sin \; x$ for LQC. 

Considering the similarities between the general relativity
equations for stars and for cosmology, and considering the
structure of the quantum effective equations for cosmology
obtained using LQC, we now propose to analogously modify the
general relativity equations (\ref{el}) and (\ref{le}) for stars
also. Consider the variables $m^\alpha$ where
\[
\alpha = (0, i, a) \; \; , \; \; \; m^\alpha = (m^0, m^i, m^a)
\]
with $i = 1, 2, \cdots, n_c \; , \; a = 1, 2, \cdots, n \;$, and
$m^a = \hat{m}$ for all $a \;$. In the model we propose, the
conservation equations (\ref{1*Il}) remain the same but
equations (\ref{el}) and (\ref{le}), which is equivalent to
(\ref{lll}), are modified. The model is specified by one
arbitrary function $f(x)$ with the only requirement that $f(x)
\to x $ in the limit $x \to 0 \;$. The general relativity
equations follow for $f(x) = x \;$. In terms of the functions
$f^\alpha, \; g_\alpha$, and $X_\alpha$ defined by
\begin{equation}\label{afgx} 
f^\alpha = f(m^\alpha) \; \; , \; \; \;
g_\alpha = \frac{d \; f^\alpha} {d m^\alpha} \; \; , \; \; \;
X_\alpha = g_\alpha \sum_\beta G_{\alpha \beta} f^\beta \; \; ,
\end{equation}
we propose that the modified equations for the $D = (n_c + n +
2)$ dimensional stars be given by
\begin{eqnarray}
\sum_{\alpha \beta} G_{\alpha \beta} \; f^\alpha \; f^\beta
& = & 2 \; l_{qm}^2 \kappa^2 \; \tilde{\Pi} \label{emod} \\
& & \nonumber \\
(m^\alpha)_l \; + \; \sum_\beta \frac {(m^\alpha - m^\beta) \;
X_\beta} {(D - 2) \; l_{qm}} & = & l_{qm} \kappa^2 \; \sum_\beta
G^{\alpha \beta} \; \left( - \tilde{\Pi}
+ \tilde{p}_\beta \right) \label{ml} \\
& & \nonumber \\ 
l_{qm} \; \lambda^\alpha_l & = &
\sum_\beta G^{\alpha \beta} X_\beta \nonumber \\
& & \nonumber \\ 
\longleftrightarrow \; \; \; X_\alpha \; = \; l_{qm} \; Y_\alpha
& = & l_{qm} \; \sum_\beta G_{\alpha \beta} \; \lambda^\beta_l
\label{lmod}
\end{eqnarray}
where $\tilde{\Pi}$ and $\tilde{p}_\alpha$ are given by
equations (\ref{1*Il}) and the constant $l_{qm}$ is a length
parameter which may characterise, a la LQC, the quantum of the
$(D - 2)$ dimensional area : $l_{qm}^{D - 2} = {\cal O}(1) \;
\kappa^2 \;$. Note that, upon using equation (\ref{lmod}) for
$\lambda^\alpha_l \;$, the conservation equations in
(\ref{1*Il}) and (\ref{1*l}) may be written in terms of
$X_\alpha$ as
\begin{eqnarray}
l_{qm} \; (\Pi_{\tilde{I}})_l & = & \sum_{\alpha \beta}
G^{\alpha \beta} \; (- \Pi_{\tilde{I}} + p_{\alpha \tilde{I}})
\; X_\beta \label{1Ix} \\
& & \nonumber \\
l_{qm} \; \tilde{\Pi}_l & = & \sum_{\alpha \beta}
G^{\alpha \beta} \; (- \tilde{\Pi} + \tilde{p}_\alpha) \;
X_\beta \label{1x} \; \; .
\end{eqnarray}
Equation (\ref{1x}) also follows upon calculating $\Pi_l$ from
equation (\ref{emod}) and then using equation (\ref{afgx}) for
$X_\alpha$ and (\ref{ml}) for $(m^\alpha)_l \;$. Equivalently,
equation (\ref{emod}) may be derived as an integral of equations
(\ref{ml}) and (\ref{1x}).

Note that for any linear function
$f(x) = c \; x + c_0$ where $c$ and $c_0$ are constants, one has
\begin{equation}\label{gac}
f^\alpha = c \; m^\alpha + c_0 \; \; , \; \; \;
g_\alpha = c \; \; , \; \; \;
X_\alpha = c \; \sum_\beta G_{\alpha \beta} f^\beta 
\; \; , \; \; \;
l_{qm} \; \lambda^\alpha_l = c \; f^\alpha \; \; .
\end{equation}
The last equation above suggests that the variables $m^\alpha$
may be thought of as quantum generalisation of the fields
$\lambda^\alpha_l \;$. If $c \ne 0 \;$ then equations
(\ref{emod}) and $(\ref{ml})$ give the general relativity
equations (\ref{el}) and (\ref{le}) with $\kappa^2$ now replaced
by $c^2 \kappa^2 \;$. If $c = 0$ then $X_\alpha = 0$ which
implies that $\lambda^\alpha , \; \Pi_{\tilde{I}} \;$, and
$p_{\alpha \tilde{I}}$ will remain constants; equation
(\ref{emod}) implies that $(D - 1) (D - 2) \; c_0^2 = 2 \;
l_{qm}^2 \kappa^2 \; \tilde{\Pi} \;$; and equation (\ref{ml})
implies that $m_\alpha$ will evolve and not remain constant.

The equations of state need to be known which give, for example,
$p_{\alpha \tilde{I}}$ in terms of $\Pi_{\tilde{I}} \;$. Then,
for a given function $f(x) \;$, equations (\ref{ml}) --
(\ref{1Ix}) may be solved as follows : Let the values of
$(m^\alpha, \; \Pi_{\tilde{I}})$ be given at an initial
$l_{init} \;$. Then the equations of state give $p_{\alpha
\tilde{I}}$ and equations (\ref{afgx}) give $(f^\alpha, \;
g_\alpha, \; X_\alpha)$ at $l_{init} \;$. Equations (\ref{ml})
and (\ref{1Ix}) then give the derivatives $(m^\alpha_l, \;
(\Pi_{\tilde{I}})_l)$ at $l_{init} \;$ which, in turn, give
$(m^\alpha, \; \Pi_{\tilde{I}})$ at $l_{init} \pm \delta \;$.
Repeating this process gives $(m^\alpha, \; \Pi_{\tilde{I}}), \;
p_{\alpha \tilde{I}} \;$, and $(f^\alpha, \; g_\alpha, \;
X_\alpha)$ for all $l \;$. Equation (\ref{lmod}) then gives
$\lambda^\alpha_l (l) \;$ which leads to $\lambda^\alpha (l)$
upon an integration. If the radial coordinate is chosen to be
the areal radius then $r(l) = e^{\sigma(l)} \;$, $ \; l(r) \;$
follows upon a functional inversion, and equation (\ref{lal})
gives $e^{- \lambda(l)} = e^\sigma \; \sigma_l \;$.


\vspace{4ex}

\begin{center} 

{\bf Modified equations for $(n + 2)$ dimensional stars}

\end{center}

\vspace{2ex} 

The modified equations for $(n + 2)$ dimensional stars follow
upon setting $n_c = 0 \;$ in equations (\ref{afgx}) --
(\ref{1x}). With $n^c = 0 \;$, one has $D = n + 2$ and $\alpha =
(0, \; a) \;$. Also, denote $m^\alpha, \; f^\alpha, \;
g_\alpha$, and $X_\alpha$ by
\begin{eqnarray} 
(m^0 , \; f^0 , \; g_0 , \; X_0) & = &
(m , \; f , \; g , \; X) \label{m} \\
& & \nonumber \\
(m^a , \; f^a , \; g_a , \; X_a) & = &
(\hat{m} , \; \hat{f} , \; \hat{g} , \; \hat{X}) \label{mhat}
\end{eqnarray}
for all $a \;$. Equation for $X_\alpha$ in (\ref{afgx}) gives $X
= n g \hat{f}$ and $\hat{X} = \hat{g} \; (f + (n - 1) \hat{f})
\;$. Then, using the expressions
\[
\sum_\beta G^{\alpha \beta} \; X_\beta \; = \;
\hat{X} - X_\alpha + \frac {X} {n} 
\]
and
\[
\sum_\beta G^{\alpha \beta} \; \left( - \tilde{\Pi}
+ \tilde{p}_\beta \right) \; = \; \tilde{p} - \tilde{p}_\alpha
- \; \frac {\tilde{\Pi} + \tilde{\rho}} {n} \; \; ,
\]
the modified equations (\ref{emod}) -- (\ref{lmod}) are given by
\begin{eqnarray} 
2 \; n \; f \hat{f} + n (n - 1) \; \hat{f} ^2 & = &
2 \; l_{qm}^2 \kappa^2 \; \tilde{\Pi} \label{ncemod} \\
& & \nonumber \\ 
n \; l_{qm} \; \hat{m}_l \; + \; (\hat{m} - m) \; X & = & - \;
l^2_{qm} \kappa^2 \; \left( \tilde{\Pi} + \tilde{\rho} \right)
\label{ncmhatl} \\
& & \nonumber \\
l_{qm} \; m_l \; + \; (m - \hat{m}) \; \hat{X} & = &
l^2_{qm} \kappa^2 \; \left( \tilde{p} + \tilde{\rho}
- \; \frac {\tilde{\Pi} + \tilde{\rho}} {n} \right)
\label{ncml}
\end{eqnarray}
\begin{equation}\label{nclmod} 
l_{qm} \; \lambda^0_l \; = \;
\hat{g} f + (n - 1) \; (\hat{g} - g) \; \hat{f} \; \; , \; \; \;
l_{qm} \; \sigma_l \; = \; g \hat{f}  
\end{equation}
where $\tilde{\Pi}, \; \tilde{\rho}$, and $\tilde{p}$ are given
by equations (\ref{1*Il}) and the conservation equations for
$\Pi_I$ now become
\begin{equation}\label{n1I}
(\Pi_I)_l \; = \; - \; (\Pi_I + \rho_I) \; \lambda^0_l
\; - \; n \; (\Pi_I - p_I) \; \sigma_l 
\end{equation}
with $\lambda^0_l$ and $\sigma_l$ being given by equations
(\ref{nclmod}). Writing $\Pi_*, \; \rho_*$, and $p_*$ explicitly
in terms of $\sigma$ and after a little algebra, equations
(\ref{ncemod}) and (\ref{ncmhatl}) give 
\begin{equation}\label{efov}
2 f \hat{f} + (n - 1) \; \hat{f} ^2 \; = \; l_{qm}^2 \; \left(
\frac {2 \kappa^2} {n} \; \Pi \; + \; (n - 1) \; e^{- 2 \sigma}
\right) \; \; ,
\end{equation}
and
\begin{eqnarray}
& &
2 l_{qm} \; \hat{m}_l \; + \; 2 \; (\hat{m} - m) \; g \hat{f} \;
+ \; 2 \; f \hat{f} \; + \; (n - 1) \; \hat{f} ^2 \nonumber \\
& & \nonumber \\
& &
= \; l_{qm}^2 \; \left( - \; \frac {2 \kappa^2} {n} \; \rho
\; + \; (n - 1) \; e^{- 2 \sigma} \right) \; \; .  \label{e+2}
\end{eqnarray}
Let the radial coordinate be the areal radius $r(l) =
e^{\sigma(l)} \;$. Then equation (\ref{lal}) gives $e^{-
\lambda(l)} = e^\sigma \; \sigma_l$ and equations (\ref{psil})
give the $l-$derivatives in terms of $r-$derivatives. Equations
(\ref{ncml}) -- (\ref{e+2}), expressed in terms of $r$ and
$r-$derivatives if required, are now the modified $(n + 2)$
dimensional TOV equations for stars. Note that one equation, for
example equation (\ref{ncml}) for $m_l \;$, follows from the
remaining ones. Also it is easy check that when $f(x) = x \;$,
equations (\ref{efov}) and (\ref{e+2}) lead to the general
relativity equations (\ref{e2ov}) and (\ref{rkr}) given in
Appendix {\bf A}.


\vspace{4ex}

\begin{center}

{\bf 4. Features of the modified equations for stars}

\end{center}

\vspace{2ex} 

Consider the modified equations (\ref{emod}) -- (\ref{1x}) for
the $(n_c + n + 2)$ dimensional stars and (\ref{ncml}) --
(\ref{e+2}) for the $(n + 2)$ dimensional stars. Once the
equations of state are known and a function $f(x)$ is given,
they may be solved as explained after equation (\ref{gac}). Two
canonical choices for $f(x)$ are $f(x) = c x + c_0$ which, by
construction, gives general relativity equations with $\kappa^2$
replaced by $c^2 \kappa^2$ and $f(x) = sin \; x$ which appears
in LQC. See \cite{k17} for several other choices.

In LQC, an explicit solution for $f(x) = sin \; x$ turns out to
be possible for an isotropic universe whose constituent's
equation of state is given by $p = (const) \; \rho \;$. In the
present context, an $(n + 2)$ dimensional isotropic star of
constant density is perhaps a similar simplest case which we
take as an illustrative example here. Its solution in general
relativity can be obtained explicitly which we have presented in
Appendix {\bf A}, pointing out its relevant features.

However, we are unable to obtain explicit solutions for such a
star when the equations are modified and the function $f(x)$ is
non trivial. Hence, in this paper, we analyse the modified
equations from a general perspective, aiming only to understand
qualitatively the salient features of their solutions.


\vspace{4ex}

\begin{center}

{\bf Non singularity of the solutions}

\end{center}

\vspace{2ex} 

Consider the solutions of the fields with respect to $l$ as
given by the modified equations (\ref{emod}) -- (\ref{1x}). Note
that equation (\ref{*}) specifies $p_{\alpha *}$ in terms of
$\Pi_*$ and that $2 \kappa^2 \Pi_* = n (n - 1) \; e^{- 2 \sigma}
\;$. Assume that the equations of state are known which give
$p_{\alpha I}$ in terms of $\Pi_{I} \;$ for $I = 1, 2, \cdots,
{\cal N} \;$; that the function $f(x)$ and all its derivatives
are finite; and that equation (\ref{emod}) is satisfied
initially at $l_{init} \;$. Then equation (\ref{emod}) will be
satisfied for all $l$ since it can be obtained as an integral of
equations (\ref{ml}) and (\ref{1x}). And, it gives an upper
bound on $\tilde{\Pi}$ :
\begin{equation}\label{piqm}
\tilde{\Pi} \; \le \; f^2_{max} \; \Pi_{qm} 
\; \; , \; \; \; \Pi_{qm} \; = \;
\frac {(D - 1) (D - 2)} {2 \; l_{qm}^2 \kappa^2}
\end{equation}
where $f_{max}$ is the maximum value of $f(x) \;$.

With the equations of state giving $p_{\alpha \tilde{I}}$ in
terms of $\Pi_{\tilde{I}} \;$, equations (\ref{ml}),
(\ref{lmod}), and (\ref{1Ix}) now give $m^\alpha_l \;$,
$\lambda^\alpha_l \;$, and $(\Pi_{\tilde{I}})_l$ as polynomials
in terms of $\Pi_{\tilde{I}}$, $\; m^\alpha$, $\; f^\alpha$ and
$g_\alpha = \frac {d \; f^\alpha} {d m^\alpha}$ where $f^\alpha
= f(m^\alpha) \;$. Differentitating these expressions repeatedly
will then give all the higher $l-$derivatives of $m^\alpha \;$,
$\lambda^\alpha \;$, and $\Pi_{\tilde{I}}$ as polynomials in
terms of $\Pi_{\tilde{I}}$, $\; m^\alpha$, $\; f^\alpha$ and the
higher derivatives of $f^\alpha$ with respect to $m^\alpha \;$.
Therefore, if $f(x)$ and all its derivatives are finite then it
follows that $\Pi_{\tilde{I}}$, $\; \lambda^\alpha_l$, all their
higher $l-$derivatives, and thereby all the spacetime curvature
invariants will also remain finite for all values of $l \;$.
Thus, the solutions given by the modified equations (\ref{emod})
-- (\ref{1x}) will be non singular if $f(x)$ and all its
derivatives are finite.

Note, in particular, the finiteness of $\Pi_* \;$. Since $\Pi_*
\propto e^{- 2 \sigma}$ and $e^\sigma$ is the areal radius, see
equation (\ref{dsl}), the finiteness of $\Pi_*$ implies that the
physical radius of the $n$ dimensional sphere will remain non
vanishing in the non singular solutions. Also, note that the
function $f(x) = x$ is not finite although all its derivatives
are finite. Hence, the corresponding general relativity
solutions will generically be singular; there is no non trivial
upper bound on $\tilde{\Pi}$ since $f_{max} = \infty$ now in
equation (\ref{piqm}); and, among other things, $\Pi_*$ may
diverge and the areal radius $e^\sigma$ may vanish.


\vspace{4ex}

\begin{center}

{\bf Modified stars of constant density : novel features}

\end{center}

\vspace{2ex} 

Now consider the modified equations (\ref{ncml}) -- (\ref{e+2})
for the $(n + 2)$ dimensional isotropic stars of constant
density $\rho \le \Pi_{qm}$ where $\Pi_{qm}$ is given in
equation (\ref{piqm}). Let $f(x) = sin \; x \;$, $\; {\cal N} =
1 \;$, and the $I-$subscripts be omitted. Then, we have
\[
\Pi = p \; \; , \; \; \;
\tilde{\Pi} = p + \Pi_* \; \; , \; \; \;
\tilde{\rho} = \rho + \rho_* \; \; , \; \; \;
\tilde{p} = p + p_* \; \; .
\]
Since $sin \; x \le 1$ and all its derivatives are finite, it
follows that $p + \Pi_* \le \Pi_{qm} \;$, that all the
$l-$derivatives of the fields are finite, and that the resulting
solutions are non singular. Hence $p$ and $\Pi_*$ are non
divergent. In particular, the central pressure $p_c < \Pi_{qm}$
and, since $2 \kappa^2 \Pi_* \propto e^{ - 2 \sigma} \;$, the
areal radius $e^\sigma > {\cal O}(1) \; l_{qm} \;$ everywhere.
If the density of the star $\rho \ll \Pi_{qm}$ then it follows
from equation (\ref{kstar}) in Appendix {\bf A} that its radius
$R_{star} \; \propto \; \rho^{- \frac {1} {2} }$ and its mass
and compactness are given given by
\begin{equation}\label{kmodstar}
M_{star} \; \propto \; \rho^{- \frac {n - 1} {2} }
\; \; , \; \; \;
K_{star} \; \simeq \; \frac {4 \; n} {(n + 1)^2} \left(
1 - \frac {(n - 1)^2 \; \rho} {n ( n + 1) \; \Pi_{qm}} \right)
\; \; .
\end{equation}
For four dimensional stars, $n = 2$ and $K_{star} \stackrel {<}
{_\sim} \frac {8} {9} \;$, close to but less than the Buchdahl
bound. If the density of the star $\rho \simeq \Pi_{qm} \;$ and
its central pressure $p_c \simeq \Pi_{qm} \;$ then it follows
from equation (\ref{kpc}) in Appendix {\bf A} that it will have
a lower compactness factor and Planckian mass and radius given
by
\begin{equation}\label{kpl}
R_{star} \; \simeq \; \left( \kappa^2 \; \Pi_{qm}
\right)^{- \frac {1} {2} } \; \simeq \; l_{qm} \; \; , \; \; \;
M_{star} 
\; \simeq \; l_{qm}^{n - 1} 
\; \; , \; \; \;
K_{star} \; \simeq \;  \frac {2 n - 1} {n^2}
\; \; .
\end{equation}

With $f(x) = sin \; x \;$, we are unable to solve the modified
equations explicitly. But it is possible to understand their
solutions qualitatively. The solution will be similar to that in
general relativity as long as $p$ and $\Pi_*$ are both $\ll
\Pi_{qm} \;$ which will be the case near the surface of the star
given by $e^\sigma = R$ and $p(R) = 0 \;$. Then $m^\alpha \ll 1
\;$ and $f^\alpha \ll 1 \;$. As one goes inside the star, using
the evolution parameter $l \;$ in the modified equations,
$m^\alpha(l)$ will evolve and $f^\alpha$ will become of ${\cal
O}(1) \;$; the pressures $p$ and $\Pi_*$ will increase
monotonicaly and become of ${\cal O} (\Pi_{qm}) \;$.

For $f(x) = x \;$, the quantities $p, \; \Pi_*$, and $f^\alpha$
would continue to increase to $\infty \;$. Then the pressure $p$
would diverge and the areal radius $e^\sigma$ would vanish. For
$f(x) = sin \; x \;$ now, the pressures $p$ and $\Pi_*$ will
remain $\le \Pi_{qm} \;$ and $e^\sigma$ will remain $> {\cal
O}(1) \; l_{qm} \;$. As $m^\alpha(l)$ evolve to and beyond
$\frac {\pi} {2} \;$, one now crosses the region of maximal
pressures, hence the minimal areal radius, and enters what may
be dubbed as an `image star'.  From then on, the quantities $p,
\; \Pi_*$, and $f^\alpha$ will begin to decrease. When $p$ and
$\Pi_*$ both become $\ll \Pi_{qm} \;$, the solution will again
be similar to that in general relativity, $p$ and $\Pi_*$ will
decrease and $e^\sigma$ will increase monotonically, and $p$
will vanish eventually. Let $e^\sigma = R_{im}$ at that point,
then $p(R_{im}) = 0 \;$. Note that the radius $R_{im}$ of the
`image star' is likely to be $\simeq R \;$, the radius of the
original star, but may not be exactly equal.

Thus, for $f(x) = sin \; x \;$, as one evolves inwards from the
surface of the star at $e^\sigma = R \;$ using the modified
equations (\ref{ncml}) -- (\ref{e+2}) and the evolution
parameter $l \;$ :

\begin{itemize}

\item

The solution in the region where the pressures $p$ and $\Pi_*$
are $\ll \Pi_{qm} \;$ will be similar to that in general
relativity. Here, $m^\alpha(l) \ll 1$ and $f^\alpha \ll 1 \;$.

\item

Near the center, the pressures $p$ and $\Pi_*$ will remain
$\stackrel {<} {_\sim} \Pi_{qm} \;$ and the areal radius
$e^\sigma$ will remain $> {\cal O}(1) \; l_{qm} \;$. Beyond the
central region of maximal pressures and minimal areal radius,
will be an `image star'. Here, $m^\alpha (l) \simeq \frac {\pi}
{2} \;$ and evolve beyond $\frac {\pi} {2} \;$, and $f^\alpha =
{\cal O}(1) \;$.

\item
 
In the `image star', the solution in the region where the
pressures $p$ and $\Pi_*$ are $\ll \Pi_{qm} \;$ will again be
similar to that in general relativity. The `image star' has a
surface where $e^\sigma = R_{im} \simeq R$ and $p(R_{im}) = 0
\;$. Here, $\pi - m^\alpha(l) \ll 1$ and $f^\alpha \ll 1 \;$.

\end{itemize}

Note that the novel features of non singular center and an
`image star', deduced here qualitatively, have a close analogy
in LQC : The evolution here of $m^\alpha(l)$ to and beyond
$\frac {\pi} {2} \;$, crossing the region of maximal pressures
and the minimal areal radius, and entering the `image star' are
strikingly similar to the analogous phenomena during the
bouncing phase of an anisotropic universe in LQC where, as one
goes back into the past, the density of the universe passes
through its non divergent maximum and the scale factors pass
through their non vanishing minima. The evolution of the `image
universe' in the far past of the bounce, where again the density
decreases and the scale factors increase, is similar to that in
general relativity.

In the LQC inspired models, which generalise the effective
equations of LQC, the function $f(x)$ is required to $\to x$ in
the limit $x \to 0 \;$ but is arbitrary otherwise. A variety of
functions can lead to different non singular cosmological
evolutions, see \cite{k17} : for example, there may be a bounce
or simply a Planckian region, the bounce may or may not be
symmetric. Similarly, the modified equations for stars proposed
here can also be studied using such functions. They will lead to
different non singular stars : for example, there may be an
`image star' or simply a region of maximal pressures, the `image
star' may or may not be similar to the original one.


\vspace{4ex}

\begin{center}

{\bf 5. Observational consequences, falsifiable predictions}

\end{center}

\vspace{2ex} 

A star of constant density is unphysical but admits explicit
solutions in general relativity and serves as an illustrative
example. It shows the divergence of central pressure and the
Buchdahl bound resulting thereby. One can also perturb such
stars, study their stability properties, and obtain insights
into more realistic stars.

When the TOV equations are modified a la LQC as proposed in this
paper, explicit solutions are not possible for a star of
constant density. Novel features arise when equations are
modified a la LQC and these features can be deduced
qualitatively. From our analysis for $f(x) = sin \; x \;$, it is
clear that the two features -- {\bf (i)} of the pressures and
densities remaining finite and bounded, and {\bf (ii)} of the
emergence of an `image star' -- will be present for realistic
stars also. These features are analogous to those present for an
anisotropic bouncing universe in LQC: the densities and
pressures remain finite and there is an `image universe' in the
past of the bounce.

Clearly, a lot of work needs to be done before one can extract
observable consequences : one needs to incorporate realistic
equations of state, rotation, perturbations which may or may not
be spherically symmetric, time dependent dynamics, and the
instabilities and the resulting collapse when the stars are
sufficiently massive. However, the modifications proposed here a
la LQC are applicable only for stars which are static and
spherically symmetric in the $(n + 1)$ dimensional transverse
space. It is not clear to us how to extend these modifications
to, for example, axially symmetric or time dependent cases.

Even with these severe limitations, the boundedness of densities
and pressures and the emergence of an `image star' for $f(x) =
sin \; x$ may lead to novel and unique features :

\begin{itemize}

\item 

Massive compact objects are non singular and will have central
regions where pressures are maximal and are of ${\cal
O}(\Pi_{qm}) \;$. Their compactness factors will be close to but
less than the Buchdahl bound. Signals from the central regions
may reach outside, very likely highly red-shifted.

\item 

There will be an `image star' beyond the central region where
pressures are of ${\cal O}(\Pi_{qm}) \;$. The areal radius in
the central region are of ${\cal O}(l_{qm}) \;$.

If signals and perturbations can traverse through the central
region of maximal pressures and minimal areal radius into the
`image star' then they may get reflected at the surface at
$R_{im}$ and traverse back to the original star. This will lead
to echoes with a unique signature of having approximately double
the delay time normally expected for a compact non singular
star. The detailed profiles of such echoes may also encode the
information about the central region of maximal pressures that
they traverse twice. If these signals can be distinguished from
those arising through multiple reflections that are invariably
present in the original star then they may lead to unique,
falsifiable, predictions.

\end{itemize} 

The `image star' is an unique consequence of the modifications
of equations a la LQC with $f(x) = sin \; x \;$. Its observable
features, mentioned above, are analogous in LQC to perturbations
evolving through a bounce region of high density, and to
receiving signals and perturbations from the universe which is
in the past of the bounce. In LQC, a signal that travels from
our universe to that in the past of the bounce may not be
naturally reflected. On the other hand, in the context of stars
and the proposed modifed equations, a signal that travels from
the original star to the `image star' will encounter the surface
of the later at $R_{im}$ which provides a natural reflection of
the signal back into the original star.


\vspace{4ex}

\begin{center}

{\bf 6. Conclusion}

\end{center}

\vspace{2ex} 

We now summarise the paper and then conclude by mentioning a
couple of topics for further studies. In this paper, we
considered general relativity equations for $(n_c + n + 2)$
dimensional stars which are static and spherically symmetric in
the $(n + 2)$ dimensional transverse space. By treating the
curvature terms of the $n$ dimensional sphere as part of the
matter sector, we rewrote the equations for the stars in a form
resembling closely the general relativity equations for
anisotropic cosmology.

We then considered the effective equations in LQC, and those in
the LQC -- inspired models. These equations contain one
arbitrary function $f(x)$ which $= x$ for general relativity and
$= sin \; x$ for LQC. Considering the similarities between the
general relativity equations for stars and for cosmology, and
considering the structure of the quantum effective equations for
cosmology obtained using LQC, we proposed modified equations for
stars which contain one arbitrary function $f(x) \;$. By
construction, the choice $f(x) = x$ gives general relativity
equations; another canonical choice is $f(x) = sin \; x$ as in
LQC. Other choices of $f(x) \;$ are also possible.

General relativity equations can be solved explicitly for an $(n
+ 2)$ dimensional star of constant density. In LQC also, where
$f(x) = sin \; x \;$, explicit solutions can be found for an
isotropic universe whose constituents have an equation of state
$p = (const) \; \rho \;$. But we are unable to find explicit
solutions to the modified equations for a star for any non
trivial function $f(x) \;$. Hence, in this paper, we analysed
these modified equations from a general perspective in order to
understand qualitatively the salient features of their
solutions.

We found that the solutions are non singular if $f(x)$ and all
its derivatives are finite : the densities, the pressures, the
curvature invariants, and the inverse of the areal radius all
remain finite and non divergent. The stars of constant density
will now have non divergent pressures at their centers.
Therefore they will be non singular, and their compactness
factors can be close to but less than the Buchdahl bound. Taking
$f(x) = sin \; x \;$, we then pointed out the resulting novel
and unique features which may be observable. One of these
features is the appearance of an `image star' beyond the central
region of maximal pressures, this being similar to the bouncing
phase of an anisotropic universe in LQC where there is an `image
universe' in the past of the bounce. 

We now conclude by mentioning a couple of topics for further
studies. There are several such topics. One topic would be to
explore whether the present modified equations can be obtained
from an LQG framework. Treating the curvature terms of the
sphere in the general relativity equations as part of the matter
sector was an important step for us in arriving at the modified
equations. Perhaps, this idea may be helpful in LQG context
also.

Can the time dependent Oppenheimer -- Snyder equations for
collapse also be similarly modified, a la LQC? Such LQC type
modifications are likely to render the densities and the
pressures finite which will thereby make the evolution non
singular. It will then be of interest to use such equations to
study the end products of stellar collapses which, in general
relativity, would have resulted in black holes.


\vspace{4ex}

\begin{center}

{\bf Appendix A : $(n + 2)$ dimensional TOV equations for stars}

\end{center}

\vspace{2ex} 

In this Appendix, we take $n_c = 0 \; , \; {\cal N} = 1 \;$,
omit the $I-$subscripts, and obtain the general relativity
solution for an $(n + 2)$ dimensional isotropic star of constant
density by setting $\Pi = p \;$ and $\rho = (const) \;$. Such a
star is unphysical but serves as an illustrative example. With
$n^c = 0 \;$, we now have
\[
\alpha = (0, \; a) \; \; , \; \; \; 
\lambda^\alpha = (\lambda^0, \; \sigma) 
\; \; , \; \; \; 
p_\alpha = (- \rho, \; p) \; \; . 
\]
Hence 
\[
\Lambda = \lambda^0 + n \sigma \; \; , \; \; \;
\sum_\beta G^{\alpha \beta} \left( \Pi + p_\beta \right)
\; = \; - p_\alpha + \frac{T}{n} 
\]
where $T = \Pi - \rho + n p \;$. Equations (\ref{e}) and
(\ref{lrr}) become
\begin{eqnarray}
2 n \; \lambda^0_r \; \sigma_r + n (n - 1) \; (\sigma_r)^2
& = & 2 \; \kappa^2 \; \Pi \; e^{2 \lambda}
+ n (n - 1) \; e^{2 \lambda - 2 \sigma} \label{ne} \\
& & \nonumber \\
n \; \sigma_{r r} + n \; (\Lambda_r - \lambda_r) \; \sigma_r
& = & \kappa^2 \; (\Pi - \rho) \; e^{2 \lambda} + n (n - 1) \;
e^{2 \lambda - 2 \sigma} \label{nsrr} \\
& & \nonumber \\ 
\lambda^0_{r r} + (\Lambda_r - \lambda_r) \; \lambda^0_r & = &
\kappa^2 \; \left( \rho + \frac {T} {n} \right) \; e^{2 \lambda}
\; \; . \label{nl0rr}
\end{eqnarray}
Let the radial coordinate $r$ be the areal radius $r = e^\sigma
\;$. Also, define a `compactness function' $K(r)$ and a `mass
function' $M(r)$ by
\begin{equation}\label{KM}
K(r) \; = \; \frac {M(r)} {r^{n - 1}} \; = \;
1 - e^{- 2 \lambda(r)} \; \; . 
\end{equation}
The $(n + 2)$ dimensional line element is now given by
\begin{equation}\label{dsn+2} 
d s^2 = - \; e^{2 \lambda^0} d t^2 + \frac {d r^2} {1 - K}
+ r^2 \; d \Omega_n^2 \; \; .  
\end{equation}
Writing $e^\sigma = r$ and after a little algebra, equations
(\ref{ne}) and (\ref{nsrr}) give
\begin{eqnarray}
2 n \; r \lambda^0_r \; e^{- 2 \lambda} & = &
2 \; \kappa^2 \; r^2 \; \Pi + n (n - 1) \; K \label{e2ov} \\
& & \nonumber \\
2 n \; r \lambda_r \; e^{- 2 \lambda} \; = \; n \; r K_r & = &
2 \; \kappa^2 \; r^2 \; \rho - n (n - 1) \; K \label{rkr}
\end{eqnarray}
from which it follows that 
\begin{equation}\label{tovmr}
n \; M_r = 2 \; \kappa^2 \; r^n \; \rho 
\; \; \; \longrightarrow \; \; \;
M = \frac {2 \; \kappa^2} {n} \; \int_0^r dr \; r^n \; \rho
\; \; . 
\end{equation}
Note that the mass function $M \propto \kappa^2 \; (mass)$ and
has dimension $(length)^{n - 1} \;$. Equations (\ref{1}) and
(\ref{e2ov}) give
\begin{eqnarray}
\Pi_r & = & - \; (\Pi + \rho) \; \lambda^0_r
- n \; (\Pi - p) \; \sigma_r \label{tovpr} \\
& & \nonumber \\
\lambda^0_r & = & \frac {2 \; \kappa^2 \; r^{n + 1} \; \Pi
+ n (n - 1) \; M} {2 n \; r \; (r^{n - 1} - M)} \; \; . 
\label{tovl0}
\end{eqnarray} 
Equations (\ref{tovmr}) -- (\ref{tovl0}) are the $(n + 2)$
dimensional TOV equations for stars. Note that equation
(\ref{nl0rr}) may be derived from them.


\vspace{2ex}

\begin{center}

{\bf Stars of constant density}

\end{center}

\vspace{2ex}

To solve equations (\ref{tovmr}) -- (\ref{tovl0}), the equations
of state need to be known. Let $\Pi = p$ and $\rho = (const)$ be
the equations of state which describe an isotropic star of
constant density. Such a star is unphysical but serves as an
illustrative example. With $\rho$ a constant, equation
(\ref{tovmr}) gives
\begin{equation}\label{mk}
M(r) = \frac {2 \kappa^2 \; \rho \; r^{n + 1}} {n (n + 1)}
\; \; , \; \; \; K(r) = \frac {M(r)} {r^{n - 1}} =
\frac {2 \kappa^2 \; \rho \; r^2} {n (n + 1)} \; \; .
\end{equation}
With $\Pi = p \;$, equations (\ref{tovpr}) and (\ref{tovl0})
give
\begin{equation}\label{pr}
p_r = - \; \kappa^2 \; r \;
\frac {(p + \rho) \; \left( (n + 1) p + (n - 1) \rho \right)}
{n (n + 1) - 2 \kappa^2 \; \rho \; r^2} 
\end{equation}
which can be integrated. Taking into account that $p(R) = 0$
defines the radius $R$ of the star, it follows after an
integration that 
\begin{equation}\label{p(r)}
\frac {(n + 1) p + (n - 1) \rho} {(n - 1) \; (p + \rho)} \; = \;
\sqrt {\frac {1 - K(r)} {1 - K(R)}} 
\end{equation}
which implies that the pressure $p(r)$ is given by
\begin{equation}\label{p}
\frac {p} {\rho} = \frac {(n - 1) \;
\left(\sqrt {1 - K(r)} - \sqrt {1 - K(R)} \right)}
{(n + 1) \sqrt {1 - K(R)} - (n - 1) \sqrt {1 - K(r)}} \; \; .
\end{equation} 
Also, since $\rho$ is constant and $\Pi = p \;$, equation
(\ref{tovpr}) gives
\begin{equation}\label{l0p}
e^{2 \lambda^0} = \frac {const} {(\rho + p)^2} = 
\left( \frac {(n + 1) \sqrt {1 - K(R)} - (n - 1) \sqrt {1 - K(r)}}
{2} \right)^2 
\end{equation}
where the $const$ is fixed by requiring that $e^{2 \lambda^0}
\to (1 - K(R))$ as $r \to R$ because $e^{2 \lambda^0} = e^{- 2
\lambda} = 1 - \frac {M(R)}{r^{n - 1}}$ for $r > R \;$.

Note that $K(R)$ is the compactness factor for the star which is
defined as the ratio of its mass to the mass of a black hole of
same radius. Let the star be characterised by its constant
density $\rho$ and the pressure $p_c \;$ at its center. Then it
follows from the above expressions that the star's compactness
$K_{star} \;$, radius $R_{star} \;$, and mass $M_{star}$ are
given in terms of $p_c$ and $\rho$ by
\begin{equation}\label{KR}
K_{star} \; = \; \frac {M_{star}} {R^{n - 1}_{star}} \; = \;
\frac {2 \kappa^2 \; \rho \; R^2_{star}} {n (n + 1)} \; = \;
\frac {4 p_c \; (n p_c + (n - 1) \rho)}
{ ((n + 1) p_c + (n - 1) \rho)^2}
\end{equation}
where the third equality gives $R_{star}$ and the second
equality then gives $M_{star}$ in terms of $p_c$ and $\rho \;$.
Thus, for $p_c \gg \rho \;$, one has
$R_{star} \propto  \rho^{- \frac {1} {2} }$ and 
\begin{equation}\label{kstar}
M_{star} \; \propto \; \rho^{- \frac {n - 1} {2} }
\; \; , \; \; \;
K_{star} \; \simeq \; \frac {4 \; n} {(n + 1)^2} \left(
1 - \frac {(n - 1)^2 \; \rho} {n ( n + 1) \; p_c} \right)
\; \; .
\end{equation}
Note that $K_{star} \simeq \frac {8} {9}$ for $n = 2$ which is
the Buchdahl bound in four dimensions. If the density $\rho
\stackrel {<} {_\sim} p_c$ then one has
\begin{equation}\label{kpc}
R_{star} \; \simeq \; \left( \kappa^2 \; p_c
\right)^{- \frac {1} {2} } \; \; , \; \; \;
M_{star} \; \simeq \; \left( \kappa^2 \; p_c
\right)^{- \frac {n - 1} {2} } \; \; , \; \; \;
K_{star} \; \simeq \;  \frac {2 n - 1} {n^2}
\; \; .
\end{equation} 

\newpage 

\vspace{4ex}

\begin{center}

{\bf Appendix B : LQC -- inspired models}

\end{center}

\vspace{2ex}

In this Appendix, for ease of reference, we write down the
equations which describe anisotropic cosmology, first in general
relativity and then in the LQC -- inspired models. Let the space
be $d = D - 1$ dimensional and toroidal with $d \ge 3 \;$. Let
$x^M = (t, x^i)$ be the spacetime coordinates where $i = 1, 2,
\cdots, d \;$, the fields depend on $t$ only, and the line
element $d s$ be given by
\begin{equation}\label{dsc}
d s^2 = - \; d t^2 + \sum_i e^{2 \lambda^i} (d x^i)^2 \; \; .
\end{equation}
Let the non vanishing components of $T^M_{\; \; N}$ be given by
$\left( T^0_{\; \; 0}, \; T^i_{\; \; i} \right) = \left( - \rho,
\; p_i \right) \;$.
Then, after a straightforward algebra, the general relativity
equations (\ref{rt}) give
\begin{eqnarray}
\sum_{i j} G_{i j} \; \lambda^i_t \; \lambda^j_t
& = & 2 \kappa^2 \; \rho \label{ce} \\ 
& & \nonumber \\
\lambda^i_{t t} + \Lambda_t \; \lambda^i_t & = & \kappa^2 \;
\sum_j G^{i j} \; (\rho - p_j) \label{ltt} \\ 
& & \nonumber \\
\rho_t + \sum_i (\rho + p_i) \; \lambda^i_t
& = & 0  \label{2} 
\end{eqnarray}
where the $t-$subscripts denote $t$ derivatives and
\[
G_{i j} \; = \; 1 - \delta_{i j} \; \; , \; \; \; G^{i j}
\; = \; \frac{1}{D - 2} - \delta^{i j} \; \; , \; \; \; \Lambda
\; = \; \sum_i \lambda^i \; \; .
\]
Note that $\sum_j G^{i j} G_{j k} = \delta^i_{\; k} \;$. Also
define $Y_i$ by
\begin{equation}\label{yi}
Y_i \; = \; \sum_j G_{i j} \; \lambda^j_t
\; \; \; \Longleftrightarrow \; \; \; 
\lambda^i_t \; = \; \sum_j G^{i j} \; Y_j 
\end{equation}
so that, using equation (\ref{ce}), equation (\ref{ltt}) for
$\lambda^i_{t t}$ may be written as
\begin{equation}\label{lce} 
\lambda^i_{t t} \; + \;
\sum_j \frac {(\lambda^i_t - \lambda^j_t) \; Y_j} {D - 2}
\; = \; - \; \kappa^2 \;
\sum_j G^{i j} \; (\rho + p_j) \; \; . 
\end{equation}
Note that differentiating both sides of equation (\ref{ce}) with
respect to $t$ and using equations (\ref{yi}) and (\ref{lce})
immediately gives the conservation equation
\[
\rho_t = - \sum_i (\rho + p_i) \; \lambda^i_t \; \; .
\]

Consider the $D$ dimensional LQC -- inspired models which we
constructed earlier by a natural, straightforward, and empirical
generalisation of the effective equations in four dimensional
LQC \cite{k16} -- \cite{k19}. These models are specified, in
what is referred to as $\bar{\mu}-$scheme, by one arbitrary
function $f(x)$ with the only requirement that $f(x) \to x $ in
the limit $x \to 0 \;$. The general relativity equations follow
for $f(x) = x$ and the LQC effective equations follow for $D =
4$ and $f(x) = sin \; x \;$.

Starting with the LQC variables in four dimensions, generalising
them empirically to $D$ dimensions, and after a long algebra,
the equations for the LQC -- inspired models may be written
concisely in terms of the variables $m^i, \; i = 1, 2, \cdots, d
\;$. In these models, the conservation equation (\ref{2})
remains the same but equations (\ref{ce}) and (\ref{lce}), which
is equivalent to (\ref{ltt}), are modified. In terms of the
functions $f^i, \; g_i$, and $X_i$ defined by
\begin{equation}\label{fgx} 
f^i = f(m^i) \; \; , \; \; \;
g_i = \frac{d \; f^i} {d m^i} \; \; , \; \; \;
X_i = g_i \sum_j G_{i j} f^j \; \; ,
\end{equation}
these modified equations in the LQC -- inspired models are given
by
\begin{eqnarray}
\sum_{i j} G_{i j} f^i f^j & = & 2 \; \gamma^2 \lambda_{qm}^2
\kappa^2 \; \rho \label{e1} \\
& & \nonumber \\
(m^i)_t \; + \; \sum_j \frac {(m^i - m^j) \; X_j} {(D - 2) \;
\gamma \lambda_{qm}} & = & - \; \gamma \lambda_{qm} \kappa^2 \;
\sum_j G^{i j} \; (\rho + p_j) \label{e2} \\
& & \nonumber \\ 
\gamma \lambda_{qm} \; \lambda^i_t & = & \sum_j G^{i j} X_j
\nonumber \\
& & \nonumber \\ 
\longleftrightarrow \; \; \; 
X_i \; = \; \gamma \lambda_{qm} \; Y_i & = &
\gamma \lambda_{qm} \; \sum_j G_{i j} \; \lambda^j_t \label{e3}
\end{eqnarray}
where the constant $\gamma$ is analogous to the Barbero --
Immirzi parameter in LQC and $\lambda_{qm}$ is a length
parameter which characterises the quantum of the $(D - 2)$
dimensional area : $\lambda_{qm}^{D - 2} = {\cal O} (1) \;
\gamma \kappa^2 \;$. Note that, upon using (\ref{e3}) for
$\lambda^i_t \;$, the conservation equation (\ref{2}) may be
written in terms of $X_i$ as
\begin{equation}\label{erhot} 
(\gamma \lambda_{qm}) \; \rho_t \; = \; - \; \sum_{i j} G^{i j}
\; (\rho + p_i) \; X_j \; \; .
\end{equation} 
Equation (\ref{erhot}) also follows upon calculating $\rho_t$
from equation (\ref{e1}) and then using equation (\ref{fgx}) for
$X_i$ and (\ref{e2}) for $(m^i)_t \;$. Equivalently, equation
(\ref{e1}) may be derived as an integral of equations (\ref{e2})
and (\ref{erhot}). Also note that for any linear function $f(x)
= c x + c_0$ where $c$ and $c_0$ are constants, one has
\begin{equation}\label{gic}
f^i \; = \; c m^i + c_0 \; \; , \; \; \;
g_i \; = \; c \; \; , \; \; \;
(\gamma \lambda_{qm}) \; \lambda^i_t \; = \; c f^i \; \; .
\end{equation} 
Equations (\ref{e1}) and (\ref{e2}) then give the general
relativity equations (\ref{ce}) and (\ref{lce}) with $\kappa^2$
now replaced by $c^2 \kappa^2 \;$. The last equation above
suggests that the variables $m^i$ may be thought of as quantum
generalisation of the fields $\lambda^i_t \;$.


\end{document}